\documentclass{ecai} 
\usepackage{latexsym}
\usepackage{amssymb}
\usepackage{amsmath}
\usepackage{amsthm}
\usepackage{booktabs}
\usepackage{enumitem}
\usepackage{graphicx}
\usepackage{color}
\usepackage{soul,color}
\usepackage[caption=false,font=footnotesize]{subfig}

\newcommand{\BibTeX}{B\kern-.05em{\sc i\kern-.025em b}\kern-.08em\TeX}

\begin{document}

%%%%%%%%%%%%%%%%%%%%%%%%%%%%%%%%%%%%%%%%%%%%%%%%%%%%%%%%%%%%%%%%%%%%%%%%

\begin{frontmatter}

%%% Use this command to specify your submission number.
%%% In doubleblind mode, it will be printed on the first page.

\paperid{123} 

%%% Use this command to specify the title of your paper.

\title{Evaluating Moderation in Online Social Network}

%%% Use this combinations of commands to specify all authors of your 
%%% paper. Use \fnms{} and \snm{} to indicate everyone's first names 
%%% and surname. This will help the publisher with indexing the 
%%% proceedings. Please use a reasonable approximation in case your 
%%% name does not neatly split into "first names" and "surname".
%%% Specifying your ORCID digital identifier is optional. 
%%% Use the \thanks{} command to indicate one or more corresponding 
%%% authors and their email address(es). If so desired, you can specify
%%% author contributions using the \footnote{} command.

\author[A]{\fnms{Letizia}~\snm{Milli}\orcid{0000-0001-5283-0477}}
\author[A]{\fnms{Laura}~\snm{Pollacci}\orcid{0000-0001-9914-1943}\thanks{Corresponding Author. Email: laura.pollacci@unipi.it}%\footnote{Equal contribution.}}
}
\author[A]{\fnms{Riccardo}~\snm{Guidotti}\orcid{0000-0002-2827-7613}} 

\address[A]{Department of Computer Science, University of Pisa, Italy}

%%% Use this environment to include an abstract of your paper.

\begin{abstract}
The spread of toxic content on online platforms presents complex challenges that call for both theoretical insight and practical tools to test intervention strategies. 
In this novel research paper, we introduce a simulation-based framework that extends the classical SEIZ (Susceptible–Exposed–Infected–Skeptic) epidemic model to capture the dynamics of toxic message propagation. 
Our simulator incorporates active moderation mechanisms through two distinct variants: a \textit{basic} moderator, which implements uniform, non-personalized interventions, and \textit{smart} moderator, which leverages user-specific psychological profiles based on Dark Triad traits to apply personalized, threshold-driven moderation. 
By varying parameter configurations, the simulator allows for systematic exploration of how different moderation strategies influence user state transitions over time. Simulation results demonstrate that while generic interventions can curb toxicity under certain conditions, profile-aware moderation proves significantly more effective in limiting both the spread and persistence of toxic behavior. 
This simulation framework offers a flexible and extensible tool for studying and designing adaptive moderation strategies in complex online social systems.

\end{abstract}

\end{frontmatter}

%%%%%%%%%%%%%%%%%%%%%%%%%%%%%%%%%%%%%%%%%%%%%%%%%%%%%%%%%%%%%%%%%%%%%%%%

\section{Introduction}

The spreading of toxic content on online social platforms poses an increasingly serious threat to digital communities, undermining user well-being, degrading the quality of discourse, and facilitating the spread of harmful ideologies. 
As toxicity becomes more prevalent and sophisticated, there is growing urgency for platforms and researchers to understand, predict, and mitigate its diffusion.
Computational modeling emerged as a useful tool in this effort, drawing on insights from social psychology, network science, and epidemiology. 
Among these approaches, compartmental epidemic models, particularly the SEIZ~\cite{bettencourt2006power} (Susceptible–Exposed–Infected–Skeptic) framework, have proven effective in capturing the spread of behaviors and information in online environments~\cite{mansal2023mathematical}. 
These models provide a structured view of user states and transitions, enabling dynamic analyses of digital contagion processes.
However, traditional SEIZ-based models overlook two critical dimensions that are essential to understanding toxicity in real-world platforms. 
First, they ignore the role of moderation, assuming either passive systems or undifferentiated control mechanisms. 
Second, they treat users as psychologically homogeneous, disregarding individual differences in susceptibility to toxicity and the likelihood of engaging in toxic behavior. 
These limitations constrain the models' realism and hinder their applicability to the design and evaluation of moderation strategies.

To address these gaps, we propose an extended SEIZ-based modeling framework that integrates both moderation dynamics and user-specific psychological profiles. 
Our framework\footnote{The code is avaiable at \url{https://github.com/LauraPollacci/SEIZ-SM}.} includes two variants. 
The first, \textit{SEIZ-BasicModerator} (\textsc{seiz-bm}), introduces a generic moderator that intervenes uniformly in response to toxic messages, without considering individual traits. 
The second, \textit{SEIZ-SmartModerator} (\textsc{seiz-sm}), enhances realism by incorporating user profiles based on the Dark Triad traits\footnote{The Dark Triad is a psychological theory that groups three socially aversive personality traits: Narcissism (excessive self-focus), Machiavellianism (manipulativeness and deceit), and Psychopathy (lack of empathy and antisocial behavior).}~\cite{paulhus2002dark}, %i.e., Narcissism, Machiavellianism, and Psychopathy,
and delivering personalized, threshold-based interventions aimed at reducing toxic behavior more effectively.
These extensions improve the behavioral fidelity of epidemic-style models in the context of toxicity diffusion. 
As a result, we developed a simulation system based on agent-based modeling that allows us to test how different moderation strategies, both uniform and personalized, affect the spread and reduction of toxic behavior.
In particular, \textsc{seiz-sm} enables us to simulate adaptive, psychologically informed interventions and to explore a wide range of scenarios through what-if and counterfactual analyses. 

Our results show that personalized, threshold-driven moderation significantly outperforms uniform approaches. 
\textsc{seiz-sm} consistently reduces the number of infected (toxic) users, promotes recovery to non-toxic states, and enhances the overall stability of the system. 
These findings highlight the value of implementing moderation strategies to user profiles and offer a quantitative foundation for developing more nuanced and effective content governance tools in Online Social Network (OSNs) platforms.

%%%%%%%%%%%%%%%%%%%%%%%%%%%%%%%%%%%%%%%%%%%%%%%%%%%%%%%%%%%%%%%%%%%%%%%%

\section{Related Work}
\label{sec:related}

The proliferation of toxic content in OSNs, including hate speech, harassment, and misinformation, poses significant risks to the health of digital communities~\cite{yousefi2024examining,sheth2022defining}. 
While advances in machine learning have led to improved systems for detecting and classifying toxic content at both the comment and user levels~\cite{survey_troll_detection,cyberbullying_cyberviolence_detection,comp_ml_dl_detection_toxic_comments}, such approaches often treat toxicity as a static property, neglecting its dynamic and propagative nature. 
Recent interdisciplinary research, particularly from computational social science, has begun to reframe toxicity through an \emph{epidemiological lens}, modeling its spread across users and communities as a contagion. 
This shift emphasizes not just the presence of toxicity, but also how it diffuses through networks and evolves over time, thereby introducing the need for dynamic, population-level simulation tools.

\subsection{Epidemiological Models for Online Behavior}

Epidemiological models have long been used to describe the dynamics of contagion processes, originally in the context of infectious diseases. 
Among them, the \emph{compartmental} approach segments a population into discrete states, such as Susceptible, Infected, and Recovered (SIR)~\cite{kermack1927contribution,wilson1945law}, and models transitions between them over time.
More nuanced models, such as SEIZ (Susceptible Exposed Infected Skeptic)~\cite{bettencourt2006power}, account for cognitive and behavioral variability. 
In SEIZ, individuals may be exposed to harmful content without immediately adopting it (\textit{E}), or may actively reject it (\textit{Z}), making it especially suited to modeling phenomena like rumor and misinformation diffusion in OSNs~\cite{jin2013epidemiological,tambuscio2015fact}. 
These models have been extended to incorporate user segmentation, e.g., gullible vs. skeptic~\cite{tambuscio2018network}, dynamic networks~\cite{mathur2020dynamic}, and intervention strategies such as fact-checking~\cite{mansal2023mathematical}.
Such models are particularly useful in capturing not only the temporal evolution of contagion but also the influence of structural properties of networks, such as degree distribution, clustering, or community segregation, on the outcome of diffusion processes. 
For example, \cite{maleki2025comparative} compared SIR and SEIZ models in the context of polarizing COVID-19 narratives, finding SEIZ to be more effective in modeling resistance and re-engagement cycles.
Despite their origins in epidemiology, these models have proven flexible and interpretable in online settings. 
They allow for the definition of domain-specific transmission parameters and can be calibrated with empirical data from platforms like Twitter to model the spread of behaviors, opinions, or false content.
Nevertheless, existing applications largely focus on misinformation or rumor propagation, with limited attention given to toxicity as a behavioral contagion. 
Further, the role of platform-driven interventions, such as moderation, has not been sufficiently integrated into the compartmental framework, leaving a gap between theoretical models and practical content governance strategies.

\subsection{Toxicity as Contagion in OSNs}
Only recently has the idea of modeling toxicity itself, rather than misinformation, through an epidemiological framework gained traction~\cite{obadimu2020developing}. 
Several studies apply compartmental models to track the spread of toxic discourse, particularly during events of heightened online polarization such as the COVID-19 pandemic~\cite{maleki2022applying} or political unrest~\cite{castiello2023using}. 
Further extending this line of research, in \cite{arisman2024modeling} the SEIRS (Susceptible–Exposed-Infectious–Recovered–Susceptible) model is used to investigate the dissemination dynamics of misinformation within a community. 
Their simulations underscored the importance of behavioral shifts and societal awareness in mitigating the effects of misinformation, emphasizing the need for interdisciplinary approaches that blend epidemiological, psychological, and sociological perspectives.
These models highlight key insights, showing that toxicity spreads more easily in homophilous or loosely connected communities and that feedback mechanisms like likes or retweets can amplify toxic expressions~\cite{jiang2023social}.
While early models adopt basic SIR-like structures, they often lack the cognitive granularity found in SEIZ-based approaches, such as the ability to distinguish between passive exposure and active rejection. 
Moreover, few existing models incorporate active moderation strategies as part of the simulation, despite moderation being central to controlling toxicity on platforms.

%%%%%%%%%%%%%%%%%%%%%%%%%%%%%%%%%%%%%%%%%%%%%%%%%%%%%%%%%%%%%%%%%%%%%%%%

\section{Proposed Moderated Epidemic Models}
While traditional compartmental models such as SI (Susceptible-Infective) and SIR assume that individuals have only two possible outcomes when encountering an infected person (i.e., becoming infected or not)~\cite{keeling2008modeling}. 
However, this binary view overlooks the complexity of social media interactions, where exposure to toxic content can result in diverse reactions such as immediate propagation, temporary deliberation or complete disengagement.

To represent this process more accurately, we adopt the SEIZ model~\cite{bettencourt2006power}.  
Originally proposed to capture rumor propagation~\cite{jin2013epidemiological,tambuscio2015fact}, the SEIZ model extends classical epidemic models by introducing two intermediate states between susceptible and infected, i.e., Exposed (\textit{E}) and Skeptic (\textit{Z}). 
Individuals transition from Susceptible (\textit{S}) to Exposed (\textit{E}) or directly to Infected (\textit{I}) upon contact with an infected individual, with respective probabilities governed by model parameters. 
The Exposed state models a latent phase where individuals are influenced but not yet actively spreading content, while the Skeptic state captures those who consciously choose not to propagate it. The SEIZ model thus enables richer diffusion dynamics than SI and SIR models by incorporating deliberate hesitation and resistance, but it remains agnostic to important platform-specific mechanisms such as moderation or user heterogeneity.

While SEIZ has proven effective in capturing several dynamics, traditional implementations of the model still overlook two key dimensions. 
First, they ignore the role of moderation, an essential component of real-world social platforms. 
In most existing models, toxicity spreads unimpeded once a user transitions to the infected state, with no corrective mechanisms such as interventions by moderators or platform policies. 
Second, users are treated as homogeneous agents, disregarding the psychological and behavioral variability that shapes how individuals produce, respond to, or resist toxic content. 
In practice, users differ significantly in their susceptibility to behavioral correction, and a one-size-fits-all intervention strategy may fail to de-escalate or even backfire~\cite{cresci2022personalized}.
To address these limitations, we propose two extensions to the SEIZ model. 
With respect to the original SEIZ model, first, we introduce the notion of a moderation mechanism, absent in the classical SEIZ formulation to reflect real-world interventions by platform authorities. 
Second, we incorporate individual-level variability through psychological profiling, allowing for differential responses to moderation. 
These extensions allow for more realistic online dynamics and enable the modeling of escalation or de-escalation effects in response to interventions, which are not captured in standard SEIZ models.

The first extension, \textit{SEIZ-BasicModerator} (\textsc{seiz-bm}), introduces a moderator who monitors posts and sends warning messages to users posting toxic content. 
This moderator is profile-agnostic, meaning that each time a user posts toxic content, the moderator sends the same generic message. 
This one-size-fits-all approach can unintentionally escalate the situation, potentially making the user more aggressive and increasing the overall volume of toxic content, particularly when it fails to account for the diverse ways users respond to interventions~\cite{cresci2022personalized}.
To overcome this limitation, we introduced additional parameters in the second version, \textit{SEIZ-SmartModerator} (\textsc{seiz-sm}), to make the model more realistic. 
In \textsc{seiz-sm}, the moderator is able to infer the users' profiles and sends personalized messages aimed at reducing aggression. 
Moreover, the moderator does not intervene immediately but waits until the same user has posted several toxic messages before sending the personalized message.

\begin{figure}[t]
    \centering
    \includegraphics[width=0.6\linewidth]{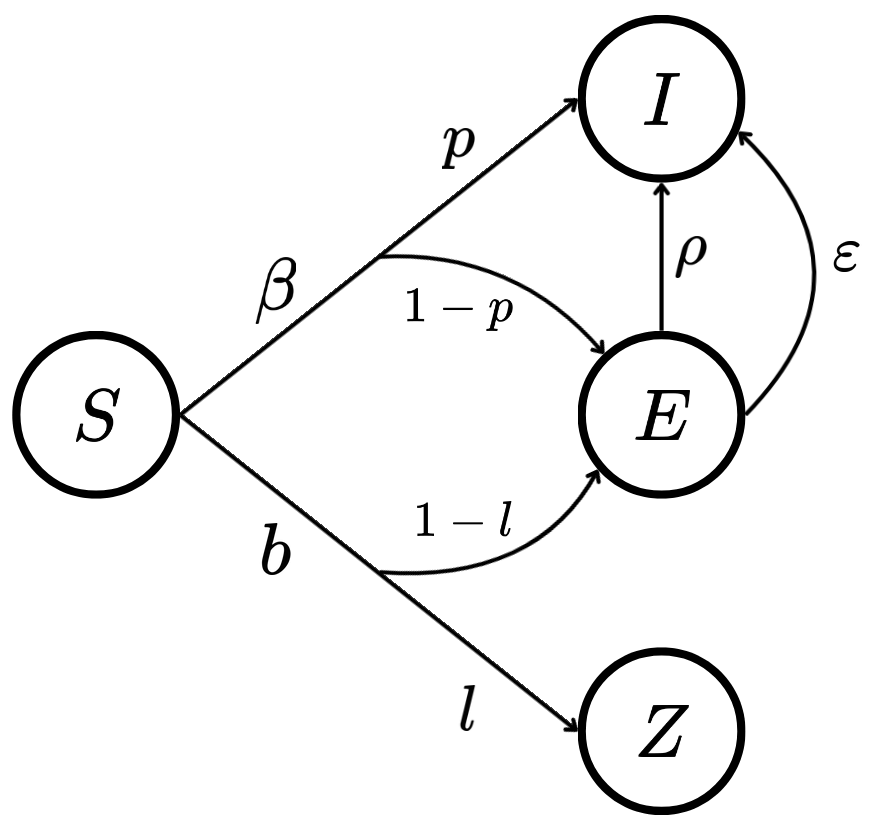}
    \caption{SEIZ model.}
    \label{fig:seiz}
\end{figure}

These models aim to bridge the gap between abstract contagion dynamics and the practical realities of content governance in online environments. 
To this end, we implement an agent-based simulation system that includes both extensions and enables systematic experimentation with intervention strategies. 
By simulating both generic and targeted moderation, the system allows us to study the effects of intervention timing, personalization, and behavioral adaptation in a dynamic, networked setting. 
This computational framework supports what-if scenario analysis and contributes to a more nuanced understanding of how moderation can influence the trajectory of toxicity spread, ultimately informing the design of more effective moderation tools and policies.
The models take into account four states:
\begin{itemize}
    \item \textit{I}: referring to an \textit{infected} user who has posted something toxic;
    \item \textit{S}: referring to a \textit{susceptible} user who is in contact, e.g., a friend or follower, of an infected user;
    \item \textit{E}: referring to an \textit{exposed} user who has come into contact with the toxic post and may potentially post a toxic message themselves;
    \item \textit{Z}: referring to a \textit{skeptical} user who has encountered the toxic post but made the conscious decision not to propagate toxicity.
\end{itemize}  
In Table \ref{tab:params}, we present the definitions of all the parameters used in the model.  
Further, Figure \ref{fig:seiz} illustrates the various states within the model, along with the rules governing the state transitions. 
When a susceptible (\textit{S}) user comes into contact with a toxic post, with a probability $\beta$, they may immediately decide to share or write a toxic message with probability $p$. 
Alternatively, they may need some time to decide what to do with a probability of $1-p$, thus entering the ``exposed'' state (\textit{E}). 
Additionally, a susceptible user might encounter a skeptical (\textit{Z}) user with probability $b$. 
In this case, two outcomes are possible: \textit{(i)} the user may also become skeptical with a probability $l$ or \textit{(ii)} the user may enter the exposed state with a probability $1-l$.

The transition from the exposed state (\textit{E}) to the infected state (\textit{I}) can occur in two different ways: \textit{(i)}  the exposed user might interact with an infected user with a probability $\rho$ leading them to post toxic content, or \textit{(ii)} without any contact, the user may independently adopt toxic behavior, transitioning to the infected state with a probability $\varepsilon$.
In the following, we provide details for the two models introduced.

\begin{table}[t]
    \centering
    \caption{Model parameters and their definitions.}
    \label{tab:params}
    \begin{tabular}{cl}
    \toprule
    \textbf{Parameter} & \textbf{Definition} \\
    \midrule
    $\beta$ & Contact rate between Susceptible and Infected (\textit{S}-\textit{I}) \\
    $b$ & Contact rate between Susceptible and Skeptic (\textit{S}-\textit{Z}) \\
    $\rho$ & Contact rate between Exposed and Infected (\textit{E}-\textit{I}) \\
    $p$ & Probability of \textit{S} becoming Infected after contact with \textit{I} \\
    $1-p$ & Probability of \textit{S} becoming Exposed after contact with \textit{I} \\
    $\varepsilon$ & Incubation rate: transition from Exposed to Infected (\textit{E} $\rightarrow$ \textit{I}) \\
    $l$ & Probability of \textit{S} becoming Skeptic after contact with \textit{Z} \\
    $1-l$ & Probability of \textit{S} becoming Exposed after contact with \textit{Z} \\
    \bottomrule
    \end{tabular}
\end{table}

\begin{figure}[t]
    \centering
    \includegraphics[width=0.7\linewidth]{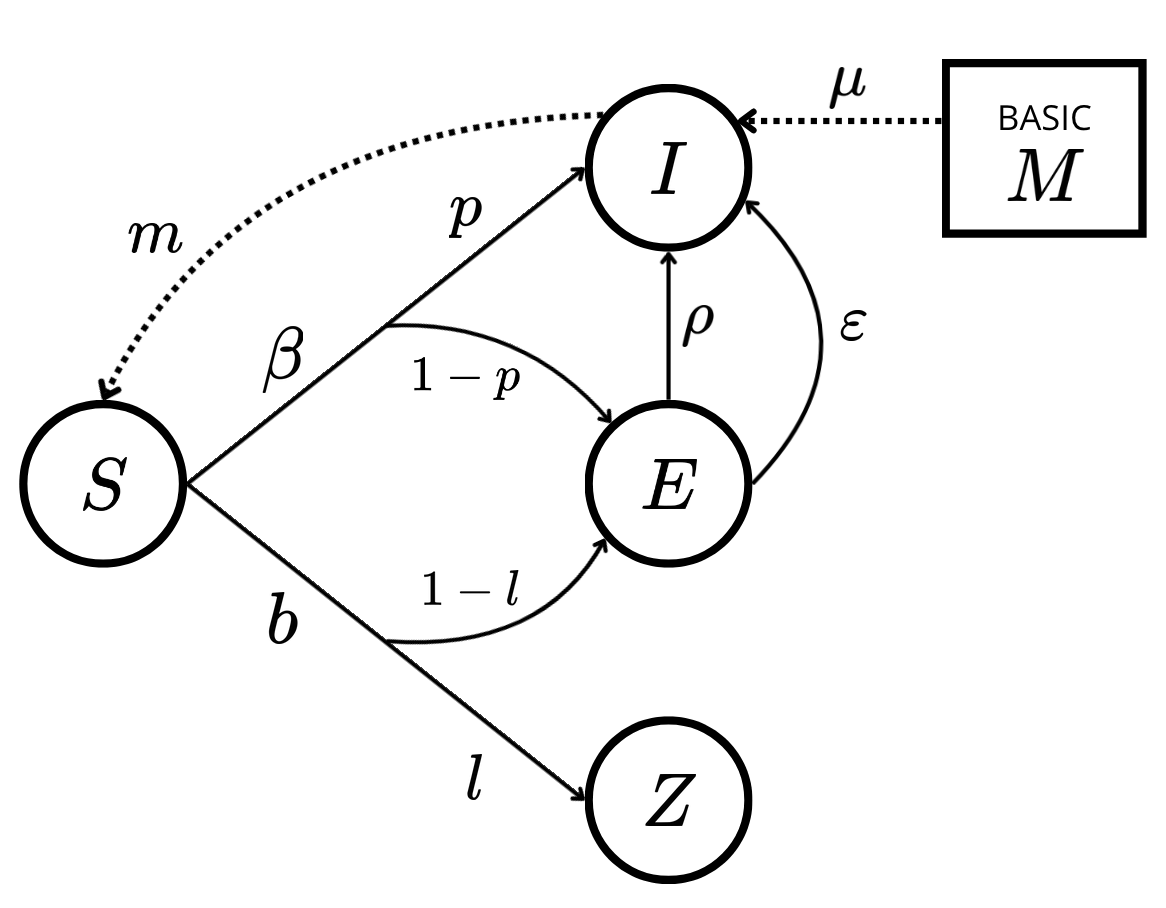}
    \caption{SEIZ-BasicModerator.}
    \label{fig:seiz_basic}
\end{figure}

\begin{figure}[t]
    \centering
    \includegraphics[width=0.7\linewidth]{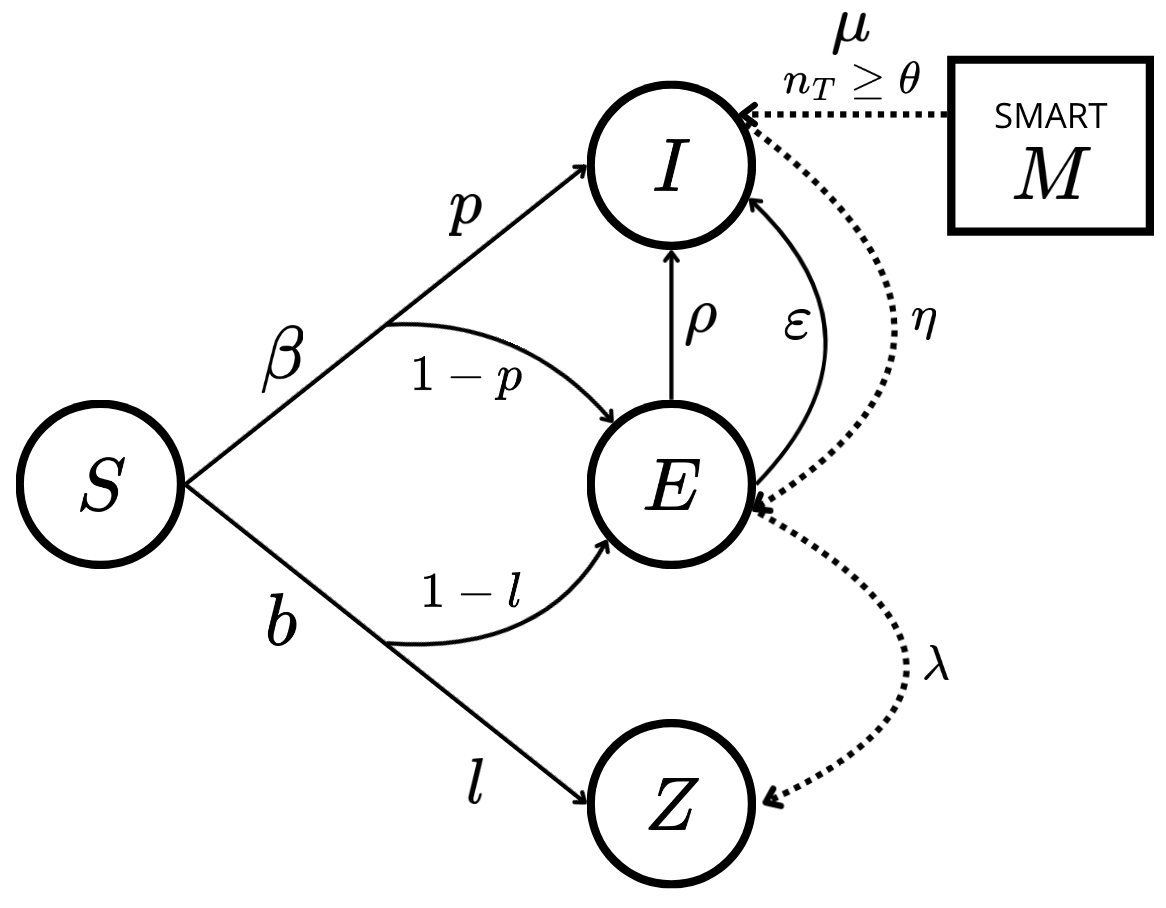}
    \caption{SEIZ-SmartModerator.}
    \label{fig:seiz_smart}
\end{figure}

\subsection{SEIZ-BasicModerator}
In \textsc{seiz-bm}, we enhanced the classical SEIZ model by introducing a moderator, that is, a user who, once a toxic message has been posted, can intervene by sending a message to the person responsible for the post. 
This model version uses the same states and transition rules as the classical SEIZ model, with an added transition rule that allows a user to move from the infected state back to the susceptible state when the moderator intervenes, as shown in Figure~\ref{fig:seiz_basic}.
As a result, two new parameters are introduced: 
\begin{itemize} 
\item $\mu$: the rate of moderator intervention; 
\item $m$: the transition rate from infected (\textit{I}) to susceptible (\textit{S}), given the moderator's intervention. 
\end{itemize}

Thus, with a probability of $\mu$, the moderator sends a message to the infected user who posted the toxic content, and with a probability of $m$, the infected user stops posting toxic messages and returns to the susceptible state. 
However, as mentioned, the moderator sends a generic message, identical for all users, without taking into account their specific profiles. 
As a result, the effectiveness of the moderation message may vary based on the user's profile. 
On one hand, it could help de-escalate the situation, or, on the other hand, if the message fails to achieve its intended effect, the user may become even more upset and continue posting toxic content.

\subsection{SEIZ-SmartModerator}
To simulate the moderator's intervention more realistically and make moderation more effective, in the \textsc{seiz-sm} model, we introduced the following new parameters:
\begin{itemize}
    \item $n$, the number of messages sent at each time step;
    \item  $\theta$, the number of toxic messages sent by an infected user before the moderator intervenes;
    \item $T$, the probability that a message is classified as toxic.
\end{itemize}

At each simulation timestep, a fixed number $n$ of messages are generated by randomly selected users. 
Only users in the infected state (\textit{I}) are capable of producing toxic content. 
Susceptible (\textit{S}), exposed (\textit{E}), and skeptic (\textit{Z}) users may also send messages, but these are always considered non-toxic and do not contribute to the spread of toxicity.
The set of toxic messages produced by infected users at each timestep determines the propagation of toxic content, as only their neighbors can be exposed and potentially become infected according to the transition probabilities defined earlier. 
This mechanism ensures that diffusion is activity-dependent with toxicity spreading only when infected users are active and selected as message senders.

Each user $u$ has a profile composed of three traits representing the Dark Triad~\cite{paulhus2002dark}, i.e., Narcissism, Machiavellianism, and Psychopathy. 
Mathematically, the user profile is defined as $p_u = [p_1, p_2, p_3]$, where $0 \leq {p_1, p_2, p_3} \leq 1$. Here, $p_1$ represents Narcissism, $p_2$ Machiavellianism, and $p_3$ Psychopathy.
These traits do not directly determine whether a user sends toxic content (only infected users can do so) but they modulate both the likelihood of becoming infected upon exposure and the probability of reverting to a non-toxic state following moderator intervention.
In particular, users with higher levels of dark traits are more prone to adopting toxic behavior and less responsive to corrective actions.

For each message sent by an infected user, a toxicity score is computed based on the sender's psychological profile: \(T_{\mathit{message}} = \tfrac{1}{3}p_1 + \tfrac{1}{3}p_2 + \tfrac{1}{3}p_3\). 
% \[
% T_{\mathit{message}} = \frac{1}{3}p_1 + \frac{1}{3}p_2 + \frac{1}{3}p_3
% \]
If $T_{\mathit{message}} \geq T$, the message is classified as ``toxic''; otherwise, it is considered ``safe''. 
When the number of toxic messages sent by an infected user exceeds the predefined threshold $\theta$, the moderator intervenes by sending a corrective \textit{personalized message}, which may induce the user to become less aggressive and transition back to the exposed state. 
In the context of our model, a \textit{personalized message} is a simulated targeted intervention whose effectiveness depends on the psychological profile of the recipient. 
It is important to note that users do not transition directly and deterministically from the infected state to the susceptible state. 
Instead, after a moderation event, they revert to the exposed (\textit{E}) state with a certain probability ($\eta$), from which they can either become re-infected or, with a certain probability $\lambda$ (that is not a subject of our study), enter the skeptic state (\textit{Z}). 
The transition from exposed to skeptic is governed by the model's probabilistic parameters and user-specific traits, and not all users will necessarily reach the skeptic state. 
Concretely, this means that the probability of transitioning from the infected to the exposed state, upon receiving a message, varies based on the user's Dark Triad scores. 
This mechanism allows the model to simulate differentiated outcomes for the same moderation action, reflecting the non-uniform behavioral response observed in real-world social platforms.
If the moderator does not intervene, perhaps because the number of toxic messages is below $\theta$, the user may escalate further and continue posting toxic content.

The extended framework captures both the complexity of human behavior and the effects of personalized moderation, providing a basis for evaluating intervention strategies through simulation, especially useful in contexts where real-world data is difficult or impossible to obtain, to study different levels of toxicity and diffusion.
%%%%%%%%%%%%%%%%%%%%%%%%%%%%%%%%%%%%%%%%%%%%%%%%%%%%%%%%%%%%%%%%%%%%%%%%

\begin{figure*}[t]
  \centering
  \subfloat[$\mu=0, m=0$]{%
    \includegraphics[width=0.33\textwidth]{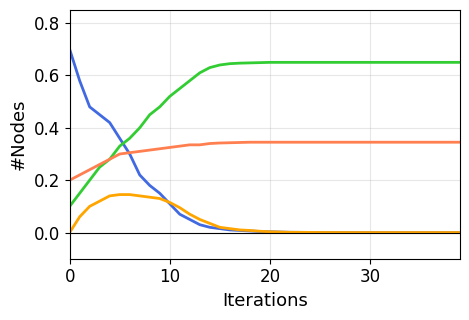}%
    \label{fig:modsub1}%
  }\hfill
  \subfloat[$\mu=0.9, m=0.3$]{%
    \includegraphics[width=0.33\textwidth]{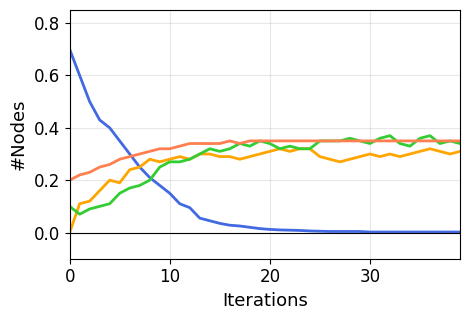}%
    \label{fig:modsub2}%
  }\hfill
  \subfloat[$\mu=0.9, m=0.9$]{%
    \includegraphics[width=0.33\textwidth]{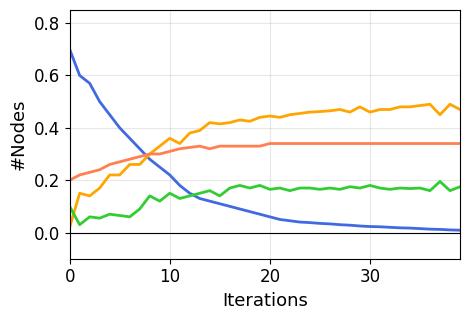}%
    \label{fig:modsub3}%
  }
  \\
  \includegraphics[width=0.5\textwidth]{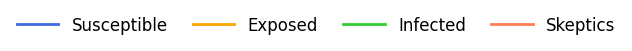}
  \caption{Simulation results per \textsc{seiz-bm}, varying the moderator intervention rate  ($\mu$) and the transition rate from Infected to Skeptic ($m$).}
  
  \label{fig:resultSEIZModerator}
\end{figure*}

\section{Experiments}
This section presents the results of simulations conducted with the two proposed SEIZ models, incorporating moderator interventions, on synthetic networks generated using the Erdős–Rényi random graph model. 
Simulations were implemented using the NDlib Python library\footnote{\url{https://ndlib.readthedocs.io/en/latest/index.html}.} \cite{rossetti2017ndlib}, which supports agent-based diffusion processes over complex networks. 
The agent-based simulation runs in discrete time steps, where user actions and state transitions are evaluated iteratively based on model parameters.

The analyses investigate how different parameter configurations affect the evolution of user states within the networked population, shedding light on the dynamics of toxic content propagation and the effectiveness of moderation strategies. 
In particular, we evaluate both generic and personalized moderation mechanisms under varying conditions, examining their impact on the temporal distribution of susceptible, exposed, infected, and skeptic users. 
The results highlight how changes in key behavioral and intervention parameters influence diffusion dynamics and demonstrate the differing impacts of basic versus personalized moderation strategies.

\subsection{Results for SEIZ-BasicModerator}
Figure \ref{fig:resultSEIZModerator} presents the results obtained with different parameter settings for the \textsc{seiz-bm} model. 
For the simulations, we kept the following parameters fixed: $\beta = 0.3$, $b = 0.3$, $\rho = 0.3$, $p = 0.3$, $l = 0.9$, and $\varepsilon = 0.3$.
On the contrary, we varied two key parameters: $\mu$, the rate at which the moderator intervenes, and $m$, the transition rate from the Infected to the Skeptic state triggered by the moderator's action.
When both variables are set to $0$ (Figure~\ref{fig:resultSEIZModerator} (a)), the model shows the standard behavior of the SEIZ model. 
However, by adjusting these parameters, for example, setting the moderator's intervention probability to $0.9$ ($\mu = 0.9$) and the probability of a successful intervention (i.e., transitioning from state I to state S) to 0.3 ($m = 0.3$), (Figure~\ref{fig:resultSEIZModerator} (b)) we observe that both the exposed and infected curves (respectively in orange and in green) stabilize around 30\% of the population, with minor fluctuations at each time step. 
In contrast, when the success rate of the moderator's intervention is high, i.e., $m = 0.9$, (Figure~\ref{fig:resultSEIZModerator} (c)) the number of susceptible users (blue) slightly increases, while the number of infected individuals (green) decreases, as expected.

\begin{figure*}[!h]
  \centering
  \subfloat[$\theta=2, n=150, T=0.5$]{%
    \includegraphics[width=0.48\textwidth]{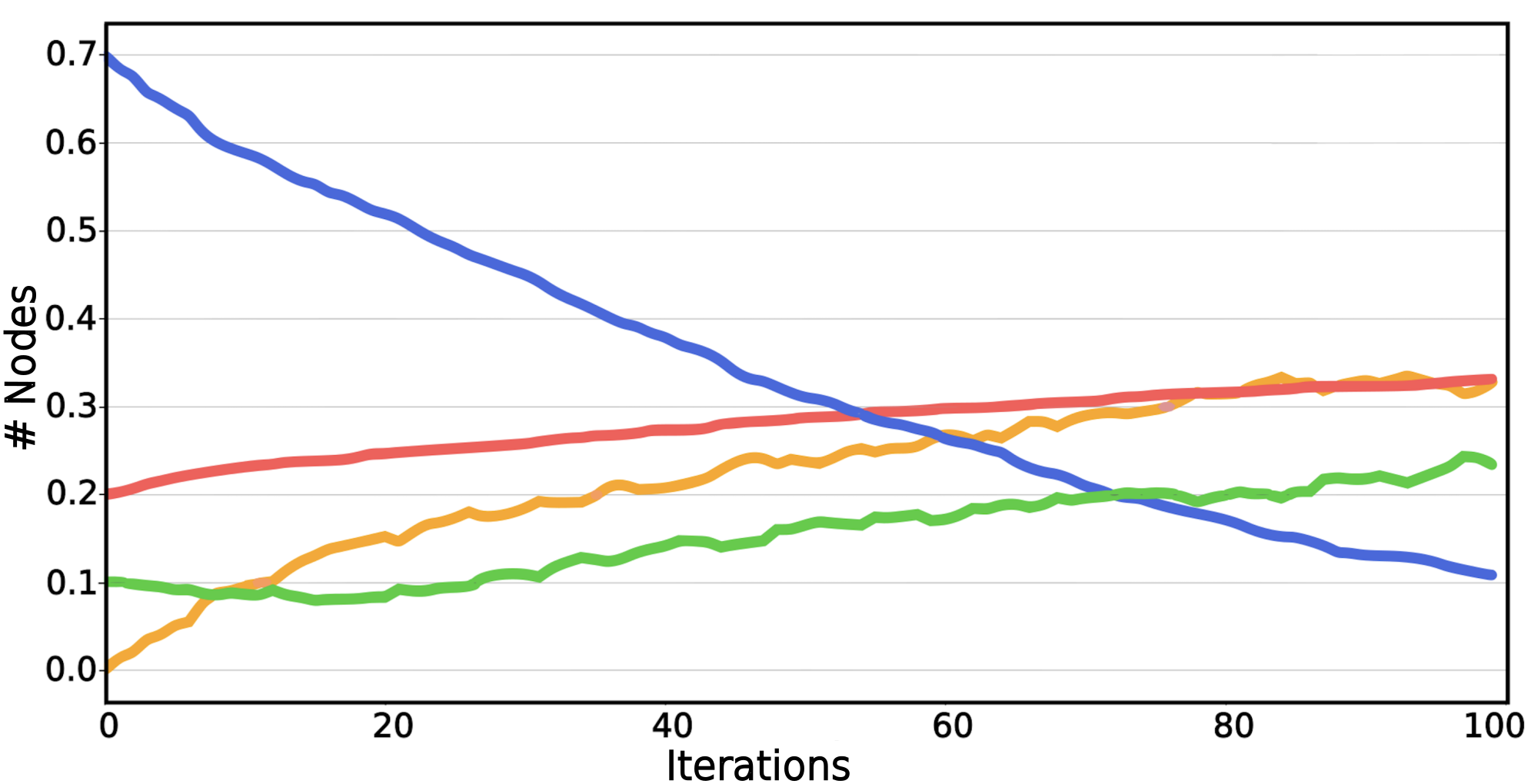}%
    \label{fig:sub1}%
  }\hfill
  \subfloat[$\theta=2, n=150, T=0.7$]{%
    \includegraphics[width=0.48\textwidth]{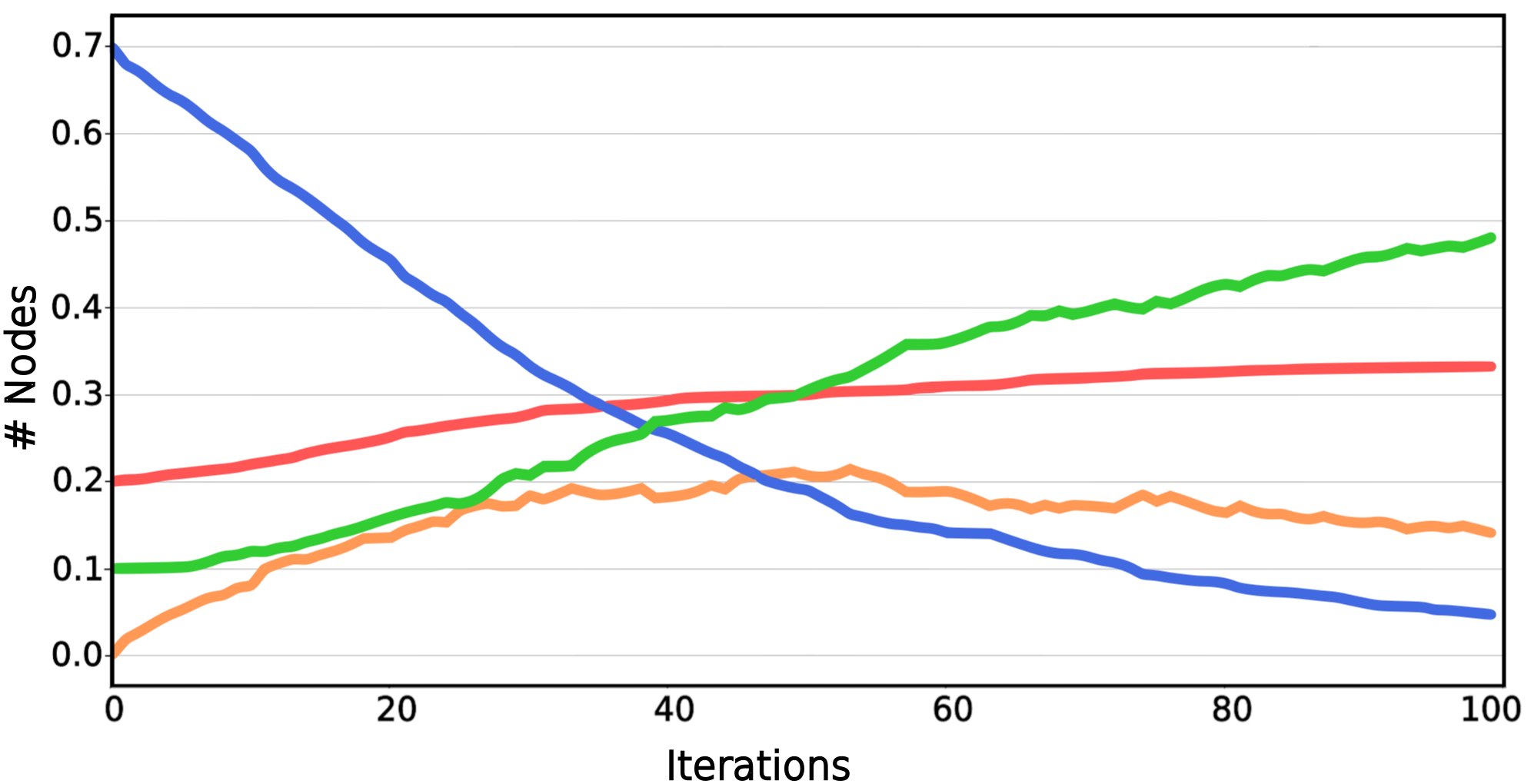}%
    \label{fig:sub2}%
  }
  \vspace{0.01em}
  \subfloat[$\theta=3, n=150, T=0.5$]{%
    \includegraphics[width=0.48\textwidth]{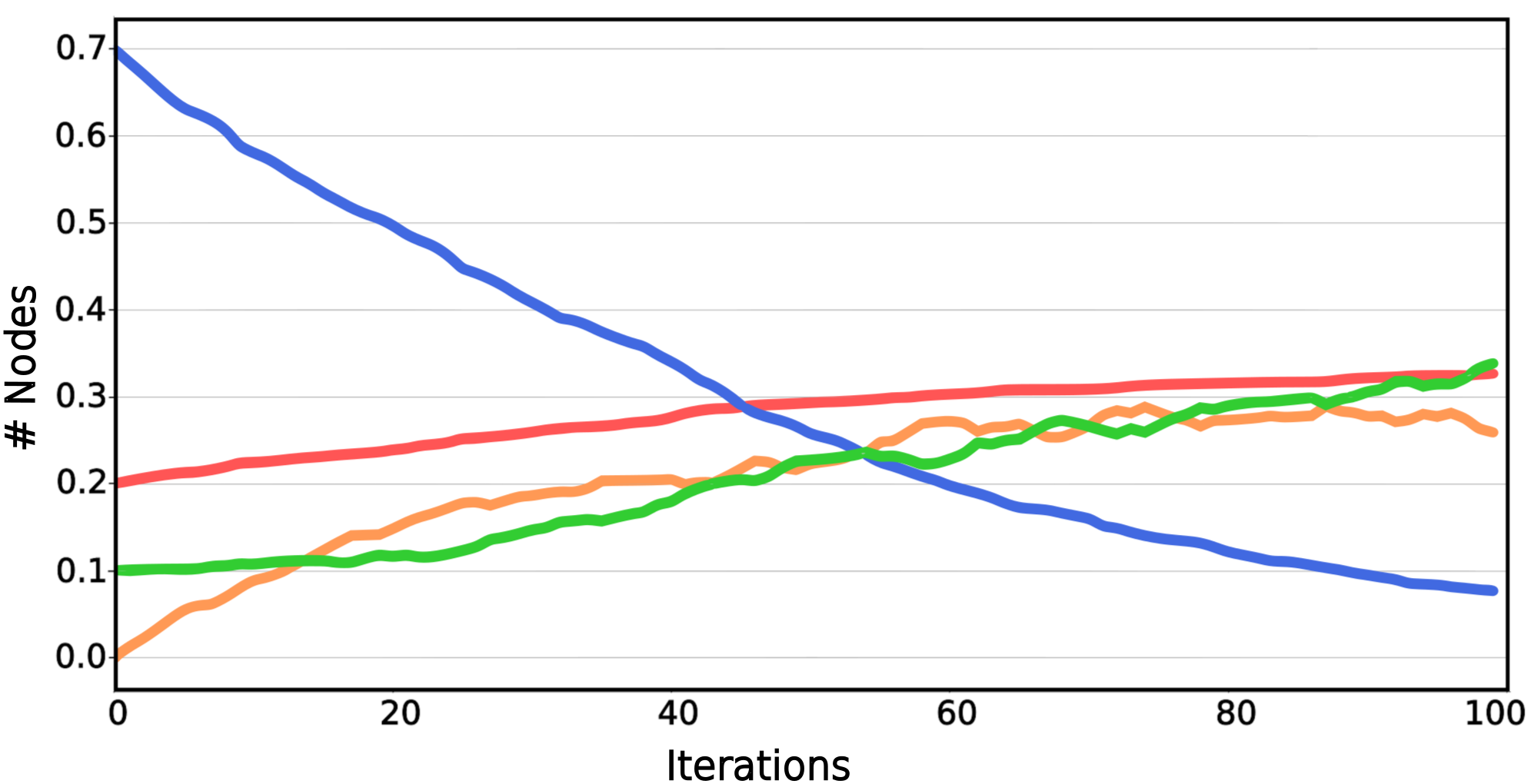}%
    \label{fig:sub3}%
  }
  \hfill
  \subfloat[$\theta=3, n=150, T=0.7$]{%
    \includegraphics[width=0.48\textwidth]{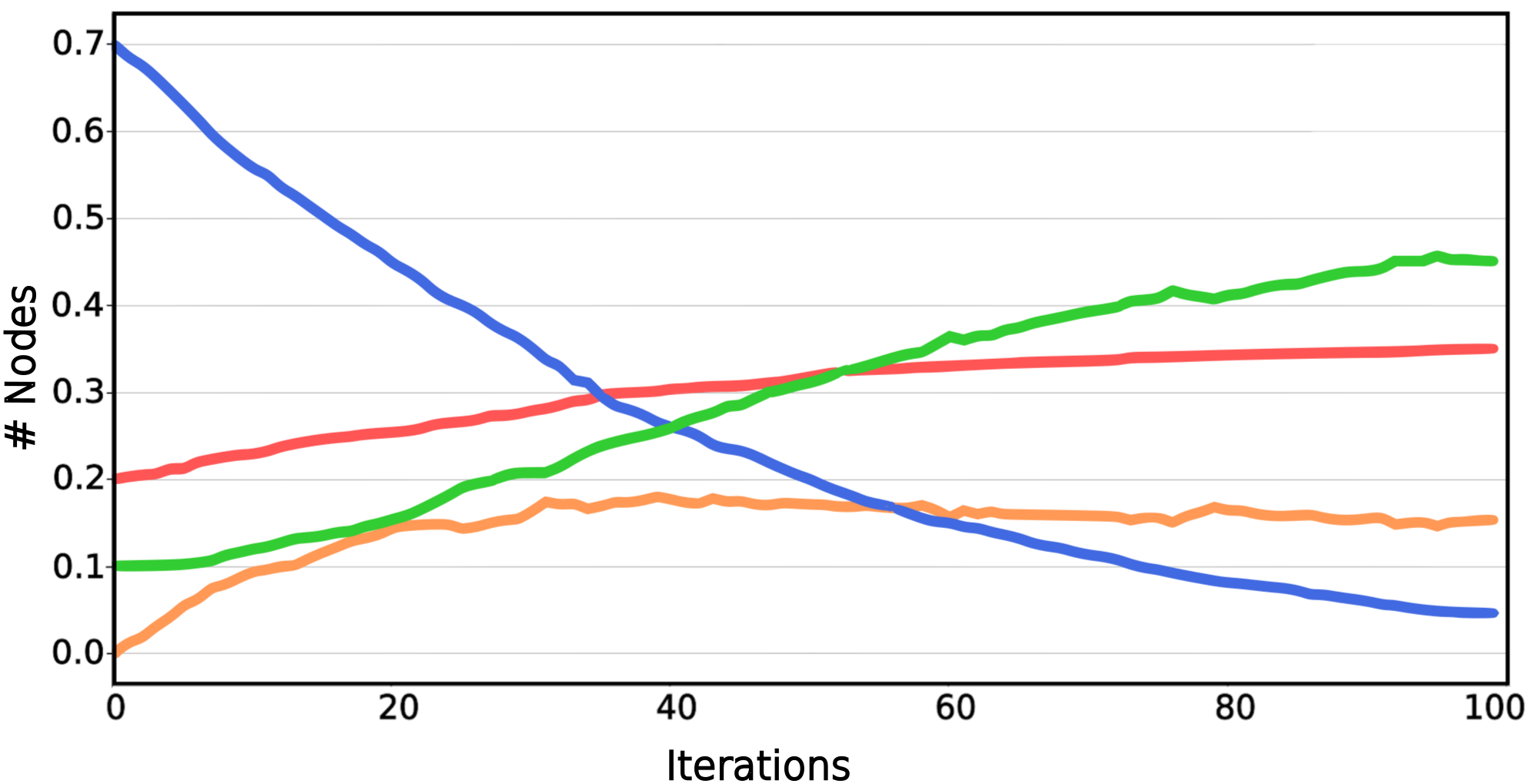}%
    \label{fig:sub4}%
  }
  \vspace{0.1em}
  \subfloat[$\theta=2, n=250, T=0.5$]{%
    \includegraphics[width=0.48\textwidth]{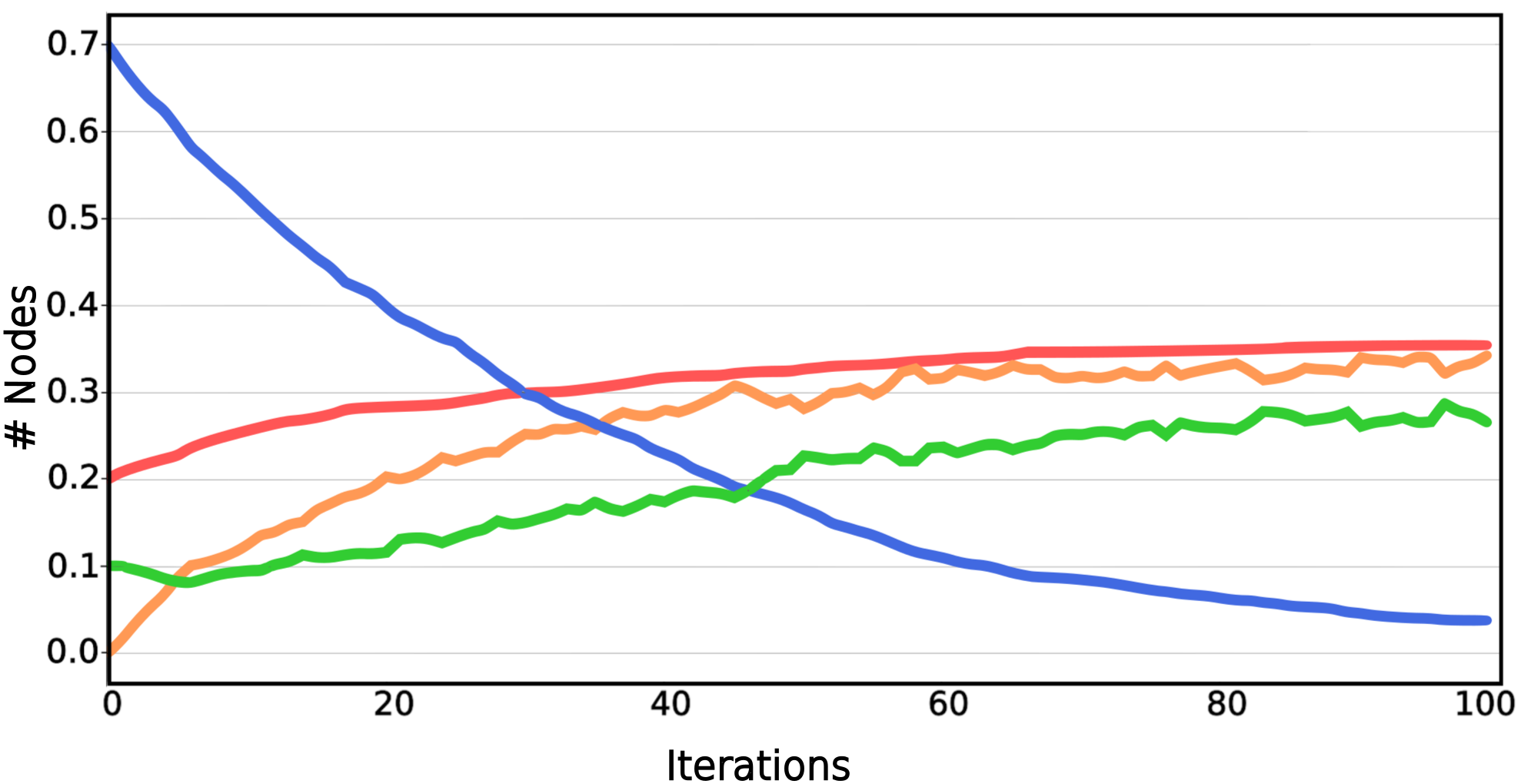}%
    \label{fig:sub5}%
  }\hfill
  \subfloat[$\theta=2, n=250, T=0.7$]{%
    \includegraphics[width=0.48\textwidth]{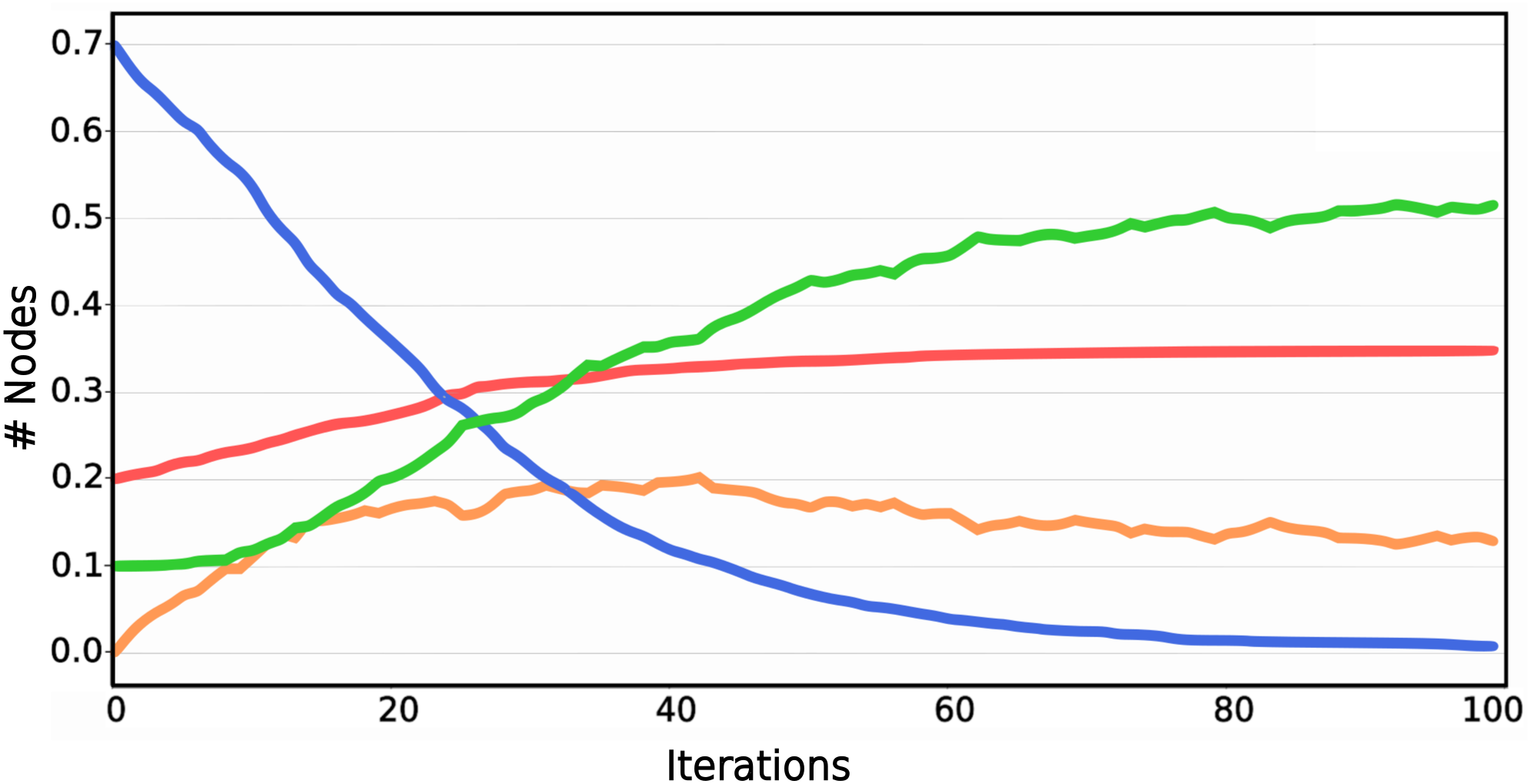}%
    \label{fig:sub6}%
  }
  \vspace{0.1em}
  \subfloat[$\theta=3, n=250, T=0.5$]{%
    \includegraphics[width=0.48\textwidth]{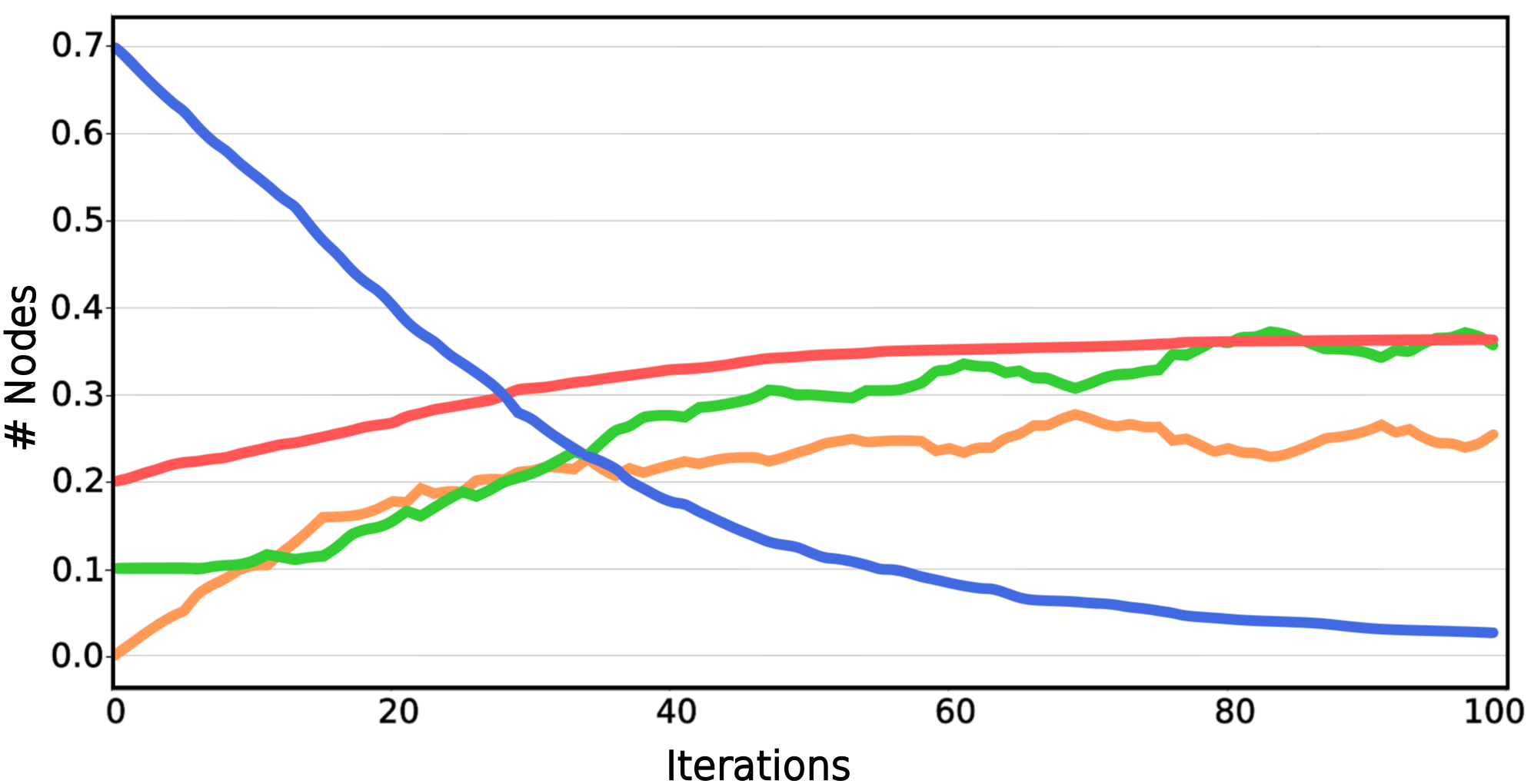}%
    \label{fig:sub7}%
  }
  \hfill
   \subfloat[$\theta=3, n=250, T=0.7$]{%
    \includegraphics[width=0.48\textwidth]{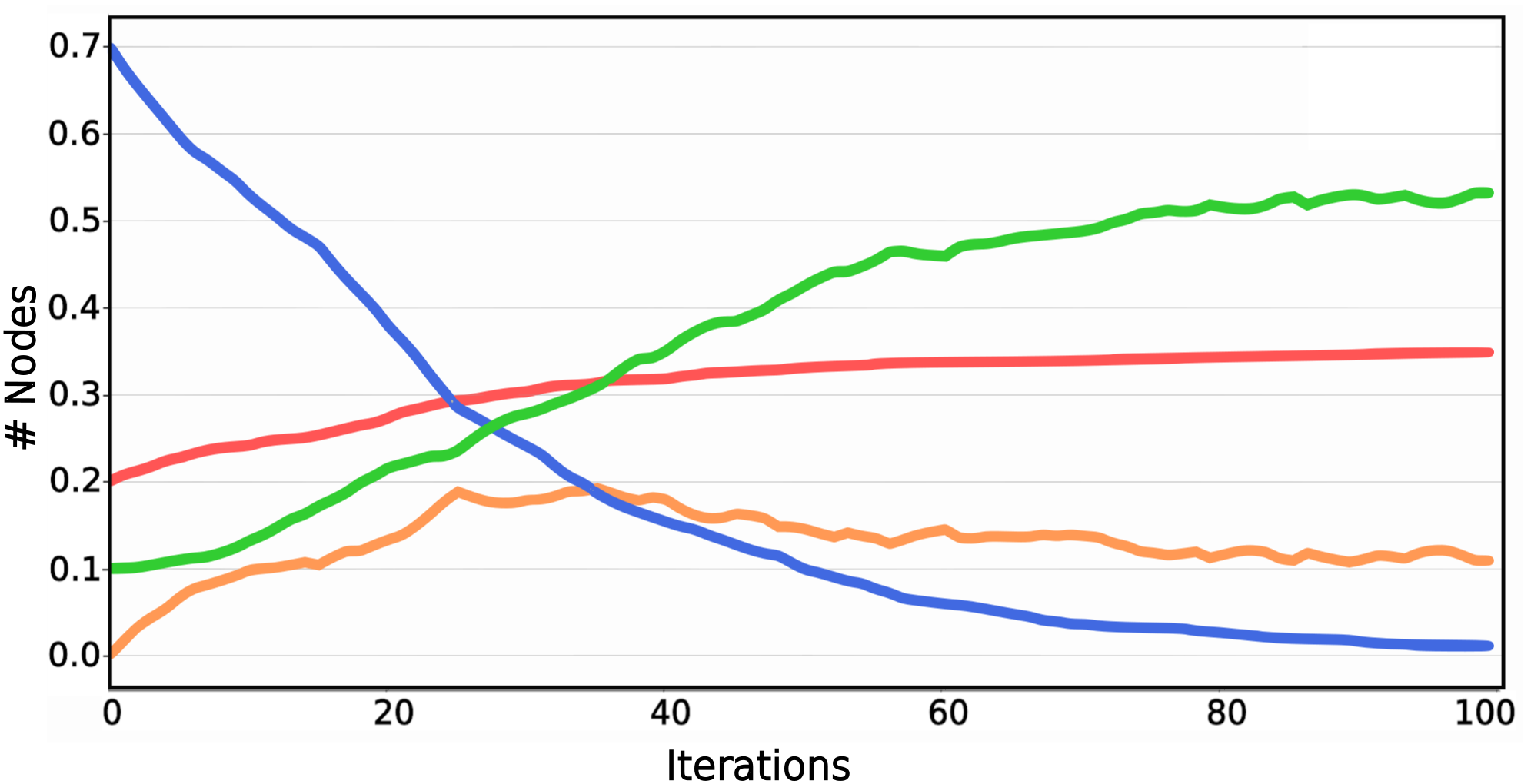}%
    \label{fig:sub8}%
  }
  \\
  \includegraphics[width=0.5\textwidth]{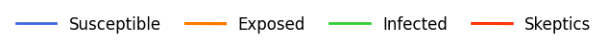}
  \caption{SEIZ‑SmartModerator simulation results under varying moderation threshold ($\theta$), message rate ($n$) e toxicity probability ($T$).}
  \label{fig:resultSEIZSmartModerator}
\end{figure*}

\subsection{Results for SEIZ-SmartModerator}

Figure \ref{fig:resultSEIZSmartModerator} presents the results obtained with different parameter configurations for the \textsc{seiz-sm} model. 
In all simulations, we kept fixed $\beta = 0.3$, $b = 0.3$, $\rho = 0.3$, $p = 0.3$, $l = 0.9$, and $\varepsilon = 0.3$.  
In contrast, we varied the number of toxic messages sent by an infected user before the moderator intervenes ($\theta$), the number of messages sent at each time step ($n$), and the probability that a message is classified as toxic (\textit{toxicity}). 
In Figure~\ref{fig:resultSEIZSmartModerator}(a), where $\theta$ is set to $2$, the number of messages per iteration is $n = 150$, and the toxicity classification threshold is relatively moderate ($0.5$), we observe a relatively balanced system evolution. 
The susceptible population (blue) decreases progressively, while the infected population (green) grows at a moderate pace. 
The skeptic population (red) remains relatively stable, and the exposed population (orange) shows a consistent upward trend. 
The early and frequent moderator intervention, enabled by the low threshold, prevents toxic behaviors from spreading excessively.
Conversely, Figure~\ref{fig:resultSEIZSmartModerator}(b) retains the same message volume and intervention threshold but increases the toxicity classification threshold to $0.7$. 
This change causes a noticeable acceleration in the growth of the infected population (green). 
Since fewer messages meet the higher toxicity threshold, fewer interventions are triggered, allowing more users to remain infected. 
The exposed population (orange) initially rises until approximately iteration 50, after which it begins to decline. 
This suggests that, while early in the simulation some users are directed toward the exposed state rather than becoming directly infected, over time the lack of prompt moderation (due to the high toxicity threshold) allows many of these users to transition into the infected class. 
Meanwhile, the skeptic population (red) remains relatively stable throughout the simulation. 
The higher toxicity threshold leads to fewer messages being flagged as toxic, delaying moderator intervention and diminishing the overall ability to maintain users in non-infected states.
Figures~\ref{fig:resultSEIZSmartModerator}(c) and~\ref{fig:resultSEIZSmartModerator}(d) further explore the impact of increasing the threshold to $3$, effectively delaying moderator intervention. 
In (c), with moderate toxicity ($0.5$) and $n = 150$, the infected population  (green) increases more and more rapidly than in (a), and the skeptics (red) grow more slowly. 
This implies that delaying intervention, even when the toxicity is moderate, reduces the system's ability to convert infected users into skeptics. 
Figure~(d), which combines a higher threshold ($3$) with high toxicity ($0.7$), represents one of the worst-performing configurations. 
The infected population (green) quickly dominates, while the number of skeptics (red), despite the high intervention threshold and toxicity, shows slow but steady increase. 
This indicates that while moderation is less effective overall, it still has a non-negligible long-term impact, albeit delayed and diminished.
The exposed state (orange) also diminishes, indicating that users are transitioning directly from susceptible to infected without moderation intercepting their behavior.
Figures~\ref{fig:resultSEIZSmartModerator}(e) through~(h) increase the message volume to $250$ while lowering the threshold to $2$.
In (e), despite the higher message rate, the low threshold and moderate toxicity ($0.5$) allow the system to maintain moderate control over infection. 
The infected population (green) grows more quickly than in (a), but the number of skeptics (red) also increases at a steady rate, indicating that early moderation still exerts a stabilizing influence.
However, in Figure~\ref{fig:resultSEIZSmartModerator}(f), increasing toxicity to $0.7$ significantly weakens the effect of the same intervention threshold. 
The system becomes overwhelmed, as the infected population (green) increases rapidly while the skeptics curve (blue) flattens. 
This suggests that even a low threshold cannot compensate for a toxicity classifier that is too lenient when the communication volume is high.
%
%Figure~\ref{fig:resultSEIZSmartModerator}(g), which uses threshold $3$ and moderate toxicity ($0.5$), shows a delayed but eventual growth in the skeptic population (red), with the infected curve (green) stabilizing over time. Here, the effect of moderation is less pronounced early on but still manages to shift user behavior later in the simulation.
In Figure~\ref{fig:resultSEIZSmartModerator}(g), the infected (green) and exposed (orange) populations show stronger fluctuations, reflecting the delayed intervention ($\theta = 3$). Despite this short-term instability, the skeptic population (red) grows and the infected curve stops rising indefinitely. This illustrates the delayed but eventual corrective influence of personalized moderation. Although interventions occur later and the system oscillates, moderation gradually shifts part of the population away from toxic states, supporting long-term containment.
Finally, Figure~\ref{fig:resultSEIZSmartModerator}(h) combines a higher threshold ($3$), high toxicity ($0.7$), and high message volume ($250$).
The infected users (green) grow quickly, the susceptible group (blue) drops sharply, and both exposed and skeptic states (orange and red) fail to gain traction. 
This configuration represents a complete breakdown of moderation effectiveness, where toxic behavior proliferates unchecked.

\subsection{Comparative Analysis and Resilience of Personalized Moderation}
The comparative analysis between the \textsc{seiz-bm} and \textsc{seiz-sm} models highlights important differences in system resilience, here intended as the system's ability to limit the spread of infected users under varying conditions. 
While both models demonstrate the capability to curb the infection dynamics when intervention is frequent and effective, the \textsc{seiz-sm} model consistently shows greater adaptability and robustness across a wider range of parameter settings.
This increased resilience stems from the personalized and context-aware nature of the \textsc{seiz-sm}. 
Unlike the \textsc{seiz-bm}, which acts uniformly across the population, the \textsc{seiz-sm} tailors its intervention based on user-level activity (e.g., message volume) and message toxicity. 
This enables earlier and more targeted mitigation actions, particularly under moderate toxicity and low intervention thresholds (e.g., Figures~\ref{fig:resultSEIZSmartModerator}(a) and~(e)), where infection is slowed while the skeptic population steadily grows.
Moreover, the \textsc{seiz-sm} demonstrates a delayed but eventual corrective influence even under less favorable conditions (e.g., Figures~\ref{fig:resultSEIZSmartModerator}(g)), suggesting that personalized moderation mechanisms allow the system to recover over time. 
In contrast, the \textsc{seiz-bm}'s effectiveness heavily depends on fixed intervention parameters and shows less adaptability to changing conditions, often resulting in higher sustained infection levels.

Thus, the personalized approach offers greater resilience by \textit{i)} enabling earlier intervention through dynamic user-level thresholds, \textit{ii) }maintaining moderation effectiveness across a broader range of message volume and toxicity levels, and \textit{iii)} promoting long-term system stabilization even when early interventions are delayed.

%%%%%%%%%%%%%%%%%%%%%%%%%%%%%%%%%%%%%%%%%%%%%%%%%%%%%%%%%%%%%%%%%%%%%%%%

\section{Discussion}

The obtained results suggest that the introduction of a moderator into the SEIZ model represents a significant advancement in modeling the dynamics of toxic content diffusion in online social environments. 

In the \textsc{seiz-bm} variant, moderator intervention is modeled as a uniform and generic response, independent of user-specific characteristics. 
While this simplification enables easier control modeling, it introduces a structural limitation: the possibility that a standardized intervention may elicit heterogeneous effects depending on the recipient's psychological profile, potentially even exacerbating toxic behavior. This observation aligns with psychological literature, which emphasizes that responses to corrective actions are mediated by individual cognitive and affective traits.
To address this, the \textsc{seiz-sm} model incorporates user-specific psychological traits modeled via the Dark Triad dimensions (Narcissism, Machiavellianism, and Psychopathy) and introduces a conditional intervention mechanism based on the number of toxic messages emitted by an infected user. 
This approach enables personalized moderation, where interventions are not only more targeted but also better timed. 
The personalization allows the system to dynamically adapt the toxicity classification threshold and respond accordingly, increasing the likelihood of reverting infected users to the exposed state, which exhibits a lower propensity for generating toxic content.

Simulation results show that moderation effectiveness  depends on the calibration of model parameters. 
For example, a lower toxicity threshold results in earlier interventions and greater containment of the infected population, whereas higher thresholds delay moderation and lead to larger infected peaks and reduced exposed populations. 
Similarly, a higher message rate ($n$) accelerates the spread dynamics, stressing the importance of prompt intervention. 
However, it does not fundamentally alter the qualitative trends, provided the moderator responds adequately. The parameter analysis shows that lower thresholds and personalized responses significantly reduce toxicity, while excessive delays in intervention diminish the moderator's efficacy.
In scenarios with high toxicity thresholds and elevated thresholds for intervention, the moderation mechanism becomes ineffective, allowing the infected population to grow unchecked while exposed and skeptic groups stagnate or decline. 
This outcome underscores the necessity of both timely and individualized moderation to maintain system stability.
Further, when moderation is delayed, the exposed population growth can create a tipping point, where many exposed users turn infected at once, reducing containment effectiveness. In real-world settings, this could be modeled by lowering moderation success when infection levels are already high.

From a modeling perspective, these results highlight the relevance of integrating both user profile information and adaptive moderation policies into diffusion models. Effective moderation cannot rely solely on frequency or volume of interventions, but must also account for behavioral profiles and the system's responsiveness to behavioral cues. 
The \textsc{seiz-sm} thus contributes a realistic and quantitatively grounded framework for simulating differentiated intervention strategies in complex social ecosystems.

The robustness of these findings is supported by representative simulation runs on which the results are based.
While the model incorporates probabilistic elements, repeated executions under identical parameter settings consistently yielded similar system dynamics, indicating stable and reproducible behavior.
Although a formal statistical analysis of variability was not conducted, the observed consistency across multiple runs provides preliminary evidence supporting the robustness of our conclusions.

%%%%%%%%%%%%%%%%%%%%%%%%%%%%%%%%%%%%%%%%%%%%%%%%%%%%%%%%%%%%%%%%%%%%%%%%

\section{Conclusions}
To support the design of more effective moderation strategies, we extended the classical SEIZ framework with two moderation-aware variants aimed at capturing how interventions influence the spread of toxicity. The \textsc{seiz-bm} model introduces a fixed, non-personalized moderation rule, while the \textsc{seiz-sm} model incorporates user-level psychological profiling via the Dark Triad and applies personalized, threshold-driven interventions to reduce aggressiveness. 
To test these models, we developed an agent-based simulation system, enabling detailed exploration of how moderation strategies unfold over time in a dynamic environment. 
Our study shows that while basic moderation may limit toxicity under certain conditions, the personalized, profile-informed approach offers greater resilience and long-term suppression of toxic behavior. 
The timing, intensity, and adaptiveness of interventions, especially relative to user traits, are key determinants of success. 
In particular, low toxicity thresholds and prompt responses prevent escalation, whereas delayed interventions can compromise moderation efforts.

This work offers a theoretical extension of epidemic-style models, and a practical simulation environment to test and compare content moderation strategies under varied conditions. 
It contributes to ongoing research at the intersection of computational social science, behavioral modeling, and moderation policy design. 
In this context, the \textsc{seiz-sm} model provides a flexible and extensible framework for exploring how individual-level characteristics and strategic interventions co-evolve within toxic content diffusion.

Despite this, the proposed models are built on two simplifying assumptions that should be acknowledged. 
First, the network structure is assumed to be static throughout the simulation. 
While common in many diffusion models, real-world OSNs are highly dynamic and users can join or leave, form or sever connections, and adjust their exposure to content through unfollowing or blocking. 
These structural changes can influence the reach and speed of toxic diffusion, as well as the effectiveness of moderation. Second, user profiles (particularly the Dark Triad traits used for personalized intervention) are modeled as fixed attributes. 
In reality, psychological traits may evolve over time, especially in response to repeated interactions or interventions. 
For example, prolonged exposure to toxic environments or successful de-escalation could gradually alter a user’s behavior. 
While incorporating dynamic profiles and evolving network structures is beyond the scope of this study, we recognize these as important directions for future work. 
Introducing temporal adaptation in both users and network topology could yield more realistic simulations and deepen understanding of long-term moderation effectiveness. Besides enriching the simulation system with evolving user traits and dynamic networks, future directions could include integrating adaptive moderation strategies, such as reinforcement learning. Validating the models against real-world data would enhance their applicability and robustness, and may inform the development of more effective and ethically grounded governance tools for managing online toxicity.

%%%%%%%%%%%%%%%%%%%%%%%%%%%%%%%%%%%%%%%%%%%%%%%%%%%%%%%%%%%%%%%%%%%%%%%%
% \section{Citations and references}

% Include full bibliographic information for everything you cite, 
% be it a book \citep{pearl2009causality}, a journal article 
% \citep{grosz1996collaborative,rumelhart1986learning,turing1950computing}, 
% a conference paper \citep{kautz1992planning}, or a preprint 
% \citep{perelman2002entropy}. The citations in the previous sentence are 
% known as \emph{parenthetical} citations, while this reference to the 
% work of \citet{turing1950computing} is an \emph{in-text} citation.
% The use of \BibTeX\ is highly recommended. 

%%%%%%%%%%%%%%%%%%%%%%%%%%%%%%%%%%%%%%%%%%%%%%%%%%%%%%%%%%%%%%%%%%%%%%%%

%%% Use this environment to include acknowledgements (optional).
%%% This will be omitted in doubleblind mode.

\begin{ack}
This work is supported by the PRIN 2022 framework project PIANO, under CUP B53D23013290006, by the Italian Project Fondo Italiano per la Scienza FIS00001966 MIMOSA; and by the NextGenerationEU – National Recovery and Resilience Plan (Piano Nazionale di Ripresa e Resilienza, PNRR) – Project: “SoBigData.it - Strengthening the Italian RI for Social Mining and Big Data Analytics” – Prot. IR0000013 – Avviso n. 3264 del 28/12/2021. 
\end{ack}

%%%%%%%%%%%%%%%%%%%%%%%%%%%%%%%%%%%%%%%%%%%%%%%%%%%%%%%%%%%%%%%%%%%%%%%%

%%% Use this command to include your bibliography file.

\bibliography{mybibfile}

@article{maleki2022applying,
  title={Applying an Epidemiological Model to Evaluate the Propagation of Toxicity related to COVID-19 on Twitter},
  author={Maleki, Maryam and Arani, Mohammad and Mead, Esther and Kready, Joseph and Agarwal, Nitin},
  year={2022}
}

@article{maleki2025comparative,
  title={A Comparative Evaluation of the SIR and SEIZ Epidemiological Models to Describe the Diffusion Characteristics of COVID-19 Polarizing Viewpoints on Online Social Networks},
  author={Maleki, Maryam and Agarwal, Nitin},
  journal={arXiv},
  year={2025}
}

@article{mansal2023mathematical,
  title={Mathematical Modeling and Optimal Control of Untrue Information: Dynamic SEIZ in Online Social Networks},
  author={Mansal, Fulgence and Faye, Ibrahima},
  journal={arXiv preprint arXiv:2309.13058},
  year={2023}
}

@article{jiang2023social,
  title={Social approval and network homophily as motivators of online toxicity},
  author={Jiang, Julie and Luceri, Luca and Walther, Joseph B and Ferrara, Emilio},
  journal={arXiv preprint arXiv:2310.07779},
  year={2023}
}

@article{sheth2022defining,
  title={Defining and Detecting Toxicity on Social Media: Context and Knowledge are Key},
  author={Sheth, Amit and Shalin, Valerie L and Kursuncu, Ugur},
  journal={Neurocomputing},
  volume={490},
  pages={312--318},
  year={2022}
}

@article{castiello2023using,
  title={Using Epidemiological Models to Predict the Spread of Information on Twitter},
  author={Castiello, Matteo and Conte, Dajana and Iscaro, Samira},
  journal={Algorithms},
  volume={16},
  number={8},
  pages={391},
  year={2023},
  publisher={MDPI}
}

@article{arisman2024modeling,
  title={Modeling the dynamics of misinformation spread on social media platforms},
  author={Arisman, Arisman and Simbolon, Hasanal Fachri Satia},
  journal={Jurnal Teknik Informatika C.I.T Medicom},
  volume={15},
  pages={297--305},
  year={2024},
  publisher={IOCSPublisher}
}

@inproceedings{jin2013epidemiological,
  title={Epidemiological modeling of news and rumors on twitter},
  author={Jin, Fang and Dougherty, Edward and Saraf, Parang and Cao, Yang and Ramakrishnan, Naren},
  booktitle={Proceedings of the 7th workshop on social network mining and analysis},
  pages={1--9},
  year={2013}
}

@article{mathur2020dynamic,
  title={Dynamic SEIZ in online social networks: epidemiological modeling of untrue information},
  author={Mathur, Akanksha and Gupta, Chandra Prakash},
  journal={International Journal of Advanced Computer Science and Applications},
  volume={11},
  number={7},
  year={2020},
  publisher={Science and Information (SAI) Organization Limited}
}

@inproceedings{tambuscio2015fact,
  title={Fact-checking effect on viral hoaxes: A model of misinformation spread in social networks},
  author={Tambuscio, Marcella and Ruffo, Giancarlo and Flammini, Alessandro and Menczer, Filippo},
  booktitle={Proceedings of the 24th international conference on World Wide Web},
  pages={977--982},
  year={2015}
}

@article{tambuscio2018network,
  title={Network segregation in a model of misinformation and fact-checking},
  author={Tambuscio, Marcella and Oliveira, Diego FM and Ciampaglia, Giovanni Luca and Ruffo, Giancarlo},
  journal={Journal of Computational Social Science},
  volume={1},
  pages={261--275},
  year={2018},
  publisher={Springer}
}

@article{survey_troll_detection,
AUTHOR = {Tomaiuolo, Michele and Lombardo, Gianfranco and Mordonini, Monica and Cagnoni, Stefano and Poggi, Agostino},
TITLE = {A Survey on Troll Detection},
JOURNAL = {Future Internet},
VOLUME = {12},
YEAR = {2020},
NUMBER = {2},
ARTICLE-NUMBER = {31},
URL = {https://www.mdpi.com/1999-5903/12/2/31},
ISSN = {1999-5903},
DOI = {10.3390/fi12020031}
}

@article{cyberbullying_cyberviolence_detection,
author={Wang, Shuwen and Zhu, Xingquan and Ding, Weiping and Yengejeh, Amir Alipour},
journal={IEEE/CAA Journal of Automatica Sinica}, 
title={Cyberbullying and Cyberviolence Detection: A Triangular User-Activity-Content View}, 
year={2022},
volume={9},
number={8},
pages={1384-1405},
doi={10.1109/JAS.2022.105740}
}

@article{paulhus2002dark,
  title={The dark triad of personality: Narcissism, Machiavellianism, and psychopathy},
  author={Paulhus, Delroy L and Williams, Kevin M},
  journal={Journal of research in personality},
  volume={36},
  number={6},
  pages={556--563},
  year={2002},
  publisher={Elsevier}
}

@article{kermack1927contribution,
  title={A contribution to the mathematical theory of epidemics},
  author={Kermack, William Ogilvy and McKendrick, Anderson G},
  journal={Proceedings of the royal society of london. Series A, Containing papers of a mathematical and physical character},
  volume={115},
  number={772},
  pages={700--721},
  year={1927},
  publisher={The Royal Society London}
}

@book{keeling2008modeling,
  title={Modeling infectious diseases in humans and animals},
  author={Keeling, Matt J and Rohani, Pejman},
  year={2008},
  publisher={Princeton university press}
}

@inproceedings{obadimu2020developing,
  title={Developing an epidemiological model to study spread of toxicity on YouTube},
  author={Obadimu, Adewale and Mead, Esther and Maleki, Maryam and Agarwal, Nitin},
  booktitle={International Conference on Social Computing, Behavioral-Cultural Modeling and Prediction and Behavior Representation in Modeling and Simulation},
  pages={266--276},
  year={2020},
  organization={Springer}
}

@inproceedings{cresci2022personalized,
  title={Personalized interventions for online moderation},
  author={Cresci, Stefano and Trujillo, Amaury and Fagni, Tiziano},
  booktitle={Proceedings of the 33rd ACM Conference on Hypertext and Social Media},
  pages={248--251},
  year={2022}
}

@inproceedings{rossetti2017ndlib,
  title={Ndlib: Studying network diffusion dynamics},
  author={Rossetti, Giulio and Milli, Letizia and Rinzivillo, Salvatore and Sirbu, Alina and Pedreschi, Dino and Giannotti, Fosca},
  booktitle={2017 IEEE international conference on data science and advanced analytics (DSAA)},
  pages={155--164},
  year={2017},
  organization={IEEE}
}

@article{bettencourt2006power,
  title={The power of a good idea: Quantitative modeling of the spread of ideas from epidemiological models},
  author={Bettencourt, Lu{\'\i}s MA and Cintr{\'o}n-Arias, Ariel and Kaiser, David I and Castillo-Ch{\'a}vez, Carlos},
  journal={Physica A: Statistical Mechanics and its Applications},
  volume={364},
  pages={513--536},
  year={2006},
  publisher={Elsevier}
}

@article{wilson1945law,
  title={The law of mass action in epidemiology},
  author={Wilson, Edwin B and Worcester, Jane},
  journal={Proceedings of the National Academy of Sciences},
  volume={31},
  number={1},
  pages={24--34},
  year={1945}
}

@inproceedings{yousefi2024examining,
  title={Examining the Impact of Toxicity on Community Structure in Social Networks},
  author={Yousefi, Niloofar and Agarwal, Nitin and DiCicco, K and Morshed, Md Samin},
  booktitle={the proceeding of the Fourteenth International Conference on Social Media Technologies, Communication, and Informatics SOTICS},
  year={2024}
}

@article{comp_ml_dl_detection_toxic_comments,
author = {Bonetti, Andrea and Martínez-Sober, Marcelino and Torres, Julio C. and Vega, Jose M. and Pellerin, Sebastien and Vila-Francés, Joan},
title = {Comparison between Machine Learning and Deep Learning Approaches for the Detection of Toxic Comments on Social Networks},
journal = {Applied Sciences},
volume = {13},
year = {2023},
number = {10},
article-number = {6038},
url = {https://www.mdpi.com/2076-3417/13/10/6038},
issn = {2076-3417},
doi = {10.3390/app13106038}
}

\end{document}